\documentstyle[12pt]{article}
 1

\begin{document}
\begin{titlepage}
\begin{flushright}
{IOA.294/93}\\
\end{flushright}
%\vspace*{5mm}
\begin{center}
{\bf  FLAVOUR CHANGING NEUTRAL CURRENTS }\\
{\bf    IN PREDICTIVE SUSY-GUTS}\\
\vspace*{1cm} {\bf G.K. Leontaris} \\
\vspace*{0.3cm}
{\it Theoretical Physics Division} \\
{\it Ioannina University} \\
{\it GR-45110 Greece} \\
\vspace*{0.3cm}
\vspace*{0.5cm}
{\bf ABSTRACT} \\
\end{center}
%\vspace*{5mm}
\noindent

Constraints from flavour changing neutral currents are discussed in the context
of predictive Grand Unified Supersymmetric Theories. Uncertainties in the
estimation of their amplitudes are minimized by using a successful mass matrix
ansatz which predicts the arbitrary parameters of the Standard Model.
Furthermore, Renormalization Group Equations are used in order to express the
sparticle masses involved, in terms of the gaugino mass $m_{1/2}$, the
universal
scalar mass $m_0$ and the trilinear coupling $A$ at the GUT scale.
Modifications in the low energy expressions for the squark masses due to a
large
top Yukawa coupling are also considered. The above results, reconciled with
experimental limits are used to put lower bounds on the parameters
$m_{1/2}$, $m_0$ and the low energy scalar masses.

\vglue 2.0cm
%\vspace*{5mm}
\noindent
\begin{flushleft}
IOA-294/93 \\
July 93
\end{flushleft}
\vfill\eject
\vfill\eject

\newpage
\setcounter{page}{1}
\vglue 0.5cm
{\bf 1. Introduction}
\vglue 0.2cm
\baselineskip=14pt
Flavour changing neutral currents(FCNC) have always provided crusial tests
for  unified models of fundamental interactions. In Supersymmetric Grand
Unified Theories (SUSY-GUTs) in particular, there are several possible
sources of flavour violations, both in the hadronic as well as in the
leptonic  sector\cite{Nil,Dug,Mas,GMas,IL,Hag}. For example, gaugino
masses, scalar masses and the induced mixing in their couplings
with the fermions, generate flavour mixing at the one loop level.
Thus, one naturally expects  that flavour violation processes are
enhanced in SUSY-GUTs.

In calculating flavour violating processes it is essential to minimize the
uncertainties of the various parameters involved. In the case of the
supersymmetric models, in addition to the uncertainties introduced
by the Standard Model parameters(experimental range of the mixing angles,
top mass etc), there are also larger uncertainties introduced by the
supersymmetric scalar masses. Recently \cite{dhr,ggl,shafi1,lv,ram},
there has been
a remarkable effort to minimize the arbitrary parameters of the Standard
Model, introducing a minimum number of inputs at the GUT scale. As a result,
 once the set of these parameters is chosen at the GUT scale, all
fermion masses, mixing angles etc, are uniquely determined at low
energies. Their evolution down to $m_W$ is governed by the
 Renormalization Group Equations(RGEs).
Flavour changing processes can be used then to test more reliably a candidate
theory or to put possible limits on the various scalar masses.

In this paper, we are going to consider the Renormalization Group
effects  on FCNCs and compare them with the experimental bounds.
In this way we may get new insight of the role of supersymmetry and try to
extract some useful bounds on the basic parameters of the theory.
As an application we are going to use a particular fermion mass matrix
ansatz\cite{lv}, which is known to give the correct predictions at
low energies. In section 2, we develop a new formalism of the ansatz
in order to minimize further the arbitrary parameters and increase the
predictability of our model. In section 3 we derive analytic expressions
of the scalar masses, especially of those affected by a large top-Yukawa
coupling. In section 4, we use flavour changing reactions to
derive  limits on the
scalar masses $m_{1/2}, m_0$ and finally in section 5 we present our
conculions.

\vglue 0.6cm
{\bf  2. Structure of Fermion Mass Matrices at
$M_{GUT}$}
 \vglue 0.4cm
The quark and lepton mass matrices  have the following
form at the GUT scale\cite{lv}
\begin{eqnarray}
M_U&=&\left(\begin{array}{ccc}0&0&x\\0&y&z\\x&z&1\end{array}
 \right)\lambda _{top}(t_0) {\upsilon
 \over\sqrt{2}}sin\beta, \label{eq:upq}\\
M_D&=&\left(\begin{array}{ccc}
0&a e^{i\phi}&0\\a e^{-i\phi}&b&0\\0&0&1\end{array} \right)
\lambda _b(t_0){\upsilon
\over\sqrt{2}} cos\beta \label{eq:downq}\\
M_E&=&\left(\begin{array}{ccc}
 0&a e^{i\phi}&0\\a e^{-i\phi}&-3b&0\\0&0&1\end{array} \right)
\lambda _{\tau}(t_0){\upsilon
\over\sqrt{2}} cos\beta \label{eq:elec}
 \end{eqnarray}
with $tan\beta \equiv {<\bar h>
\over <h>}$, $\lambda _b(t_0)=\lambda _{\tau}(t_0)$, and $\upsilon =246 GeV$.
We have taken the up-quark matrix to be symmetric, and the down quark matrix
to be hermitian. Furthermore,  the charged lepton mass matrix is related
in a standard way\cite{gj} to the down quark mass matrix at the GUT scale.

 Our ansatz has
a total of five zeros in  the quark sector (the leptonic sector is directly
related to them);  two zeros for the up and three for the down quark mass
 matrix(zeros in symmetric entries are counted only once). These zeros reduce
the number of arbitrary parameters at the GUT-scale to eight, namely
$x,y,z,a,b,\phi,\lambda _b(t_0)$, and $\lambda _{top}(t_0)$. These
non-zero entries should serve to determine the thirteen arbitary parameters
of the standard model, i.e., nine quark and lepton masses, three mixing
angles and the phase of the Cabbibo-Kobayashi-Maskawa (CKM) matrix.
Thus, as far as the charged fermion sector is concerned, we end up with five
predictions (we have discussed the predictions for the neutrino sector
in \cite{lv}). We may reduce the number of arbitary parameters
by one, if we impose more structure in the up-quark mass matrix.
We may for example relate the $(13)$, $(23)$ and $(22)$ entries, as follows
\begin{eqnarray}
y =  n z^2 {},{} x =  (n-1) z^2
\end{eqnarray}
where $n$ as will be shown, is a number in the range $n\sim (3,6)$.
Although this structure is imposed by hand \cite{lv},
on the other hand increases the predictive power of the theory,
as it reduces the number of free parameters by one.

  In order to find the  structure of the mass matrices at
the low energy scale and calculate the mass eigenstates as well as
the mixing matrices and compare them  with the experimental data, we
need to evolve them down to $m_W$, using the renormalization group
equations. The RGEs for the Yukawa couplings at the one-loop level, read
\begin{eqnarray}
16\pi^2 \frac{d}{dt} \lambda_U&=& \left( I\cdot
Tr [3 \lambda_U\lambda_U^\dagger ]  +
3 \lambda_U \lambda_U^\dagger +\lambda_D \lambda_D^\dagger
-I\cdot G_U\right) \lambda_U, \label{eq:rge1}
\\
16\pi^2 \frac{d}{dt} \lambda_D&=& \left( I\cdot
Tr [3 \lambda_D\lambda_D^\dagger +
\lambda_E \lambda_E^\dagger ]  +
3 \lambda_D \lambda_D^\dagger +\lambda_U \lambda_U^\dagger
-I \cdot  G_D\right) \lambda_D, \label{eq:rge2}
\\
16\pi^2 \frac{d}{dt} \lambda_E&=& \left( I\cdot
 Tr [ \lambda_E\lambda_E^\dagger +3
\lambda_D  \lambda_D^\dagger ]  + 3
 \lambda_E \lambda_E^\dagger -I \cdot G_E\right) \lambda_E,
\label{eq:rge3}
\end{eqnarray}
where $\lambda_\alpha$, $\alpha=U,D,E$, represent the $3$x$3$ Yukawa matrices
which are defined in terms of the mass matrices given in
Eqs(\ref{eq:upq}-\ref{eq:elec}), and $I$ is the $3$x$3$ identity matrix.
$t\equiv\ln(\mu/\mu_0)$, $\mu$ is the scale at which
the couplings are to be
determined and $\mu_0$ is the reference scale,
in our case the GUT scale. The
gauge contributions are given by
\begin{eqnarray}
G_\alpha&=&\sum_{i=1}^3 c_\alpha^i g_i^2(t),\\
g_i^2(t)&=&\frac{g_i^2(t_0)}{1- \frac{b_i}{8\pi^2} g_i^2(t_0)(t-t_0)}.
\end{eqnarray}
The $g_i$ are the three gauge coupling constants of the Standard Model and
$b_i$
are the corresponding supersymmetric beta functions. The coefficients
$c_\alpha^i$ are given by
\begin{eqnarray}
\{c_Q^i \}_{i=1,2,3} = \left\{ \frac{13}{15},3,\frac{16}{3} \right\}, \qquad
\{c_D^i \}_{i=1,2,3} = \left\{\frac{7}{15},3,\frac{16}{3} \right\},\\
\{c_U^i \}_{i=1,2,3} = \left\{ \frac{16}{15},0,\frac{16}{3}\right\}\qquad
\{c_E^i \}_{i=1,2,3} = \left\{ \frac{9}{5},3,0\right\} .
\end{eqnarray}
In what follows, we will assume that $\mu_0\equiv M_G\approx 10^{16}GeV$
and $a_i(t_0)\equiv a_G \equiv \frac{g_i^2(t_0)}{4\pi}\approx
\frac{1}{25}$. In order to evolve the Equations
(\ref{eq:rge1}-\ref{eq:rge3}) down to low energies, we also need to do
some plausible approximations. For later convenience, we define a new
parameter $tan\theta_1=z$ and
 diagonalize the up quark mass matrix at the GUT scale and obtain
the eigenstates \begin{eqnarray}
m_1(M_G)\approx -n(n-1)ptan^2\theta_1 sin^2\theta_1 \nonumber\\
m_2(M_G)\approx (n-1)ptan^2\theta_1\label{eq:Gmass}\\
m_3(M_G)\approx \frac{p}{cos^2\theta_1}\nonumber
\end{eqnarray}
with diagonalizing matrix
\begin{eqnarray}
{\it K}&=&\left(\begin{array}{ccc}
\frac{1}{{\it D_1}}&\frac{-sin\theta_1}{{\it D_2}}&\frac{(n-1)sin^2\theta_1}
{{\it D_3}}\\
\frac{sin2\theta_1}{2{\it D_1}}&\frac{1}{{\it D_2}}&\frac{sin\theta_1}
{{\it D_3}}\\
\frac{nsin^2\theta_1}{{\it D_1}}&\frac{sin\theta_1}{{\it D_2}}&
\frac{1-nsin^2\theta_1}{{\it D_3}}
\end{array} \right) \label{eq:upk}
\end{eqnarray}
where we have set $p=\lambda_t(t_0){\upsilon \over\sqrt{2}}sin\beta $ and ${\it
D_1}\approx \sqrt{ 1+sin^2\theta_1 cos^2\theta_1}$, ${\it D_2}\approx
\sqrt{1+sin^2\theta_1 }$, and  ${\it D_3}\approx \sqrt{1-(2n-1)sin^2\theta_1
}$.
In this case all other mass matrices are also rotated by the same similarity
transformation, thus
\begin{eqnarray}
\lambda_{\alpha}&\rightarrow &{\tilde\lambda}_{\alpha}=
 K^\dagger\lambda_{\alpha}K,\alpha =D,E
\end{eqnarray}
 Now, assuming that the $\lambda_{top}$ coupling is much bigger
than all other fermion Yukawa couplings, we may ignore the contributions of
the latter in the RHS of the RGEs in Eqs(\ref{eq:rge1}-\ref{eq:rge3}).
In this case all differential equations reduce to a simple uncoupled form.
Thus the top-Yukawa coupling differential equation, for example,
may be cast in the form
\begin{eqnarray}
16\pi^2 \frac{d}{dt}
\tilde\lambda_{top}&=&\tilde\lambda_{top}(6{\tilde\lambda}_{top}
^2-G_Q(t))\label{eq:topeq} \end{eqnarray}
with the solution\cite{Lopez,ggl}
\begin{eqnarray}
\tilde \lambda_{top}&=&\tilde \lambda_{top}(t_0)\xi^6\gamma_Q(t)\
\label{eq:ltop}
\end{eqnarray}
where
\begin{eqnarray}
\gamma_Q(t)&=& \prod_{j=1}^3 \left(1- \frac{b_{j,0}\alpha_{j,0}(t-t_0)}{2\pi}
\right)^{c_Q^j/2b_j},\\
\xi&=& \left( 1-\frac{6}{8\pi^2}\lambda_{top}^2(t_0)
\int_{t_0}^{t} \gamma_Q^2(t)\,dt \right)^{-1/12}\label{eq:ksi}.
\end{eqnarray}
Thus the GUT up quark mass eigenstates given in Eq(\ref{eq:Gmass})
 renormalized down to their mass scale, are given by
\begin{eqnarray}
m_u \approx \gamma_Q\xi^3 \eta_u n(n-1)ptan^2\theta_1 sin^2\theta_1\nonumber\\
m_c \approx \gamma_Q\xi^3\eta_c (n-1)ptan^2\theta_1\label{eq: upmass}\\
m_t \approx \gamma_Q\xi^6 \frac{p}{cos^2\theta_1}\nonumber
\end{eqnarray}
where $\eta_u$ , $n_c$ are taking into account the renormalization effects
from the $m_t$-scale down to the mass of the corresponding quark.
In the following we will take $\eta_u(1GeV) = n_c(1GeV)\approx 2$.
We may combine  Eqs(\ref{eq: upmass}) and give predictions
for the top quark mass and  the mixing angle $\theta_1 $
in terms of the low energy up and charm quark masses. We obtain
\begin{eqnarray}
m_t \approx
\xi^3\frac{n}{n-1}\frac{m_c^2}{m_u}\frac{\eta_u}{\eta_c^2}\label{eq: tpred}\\
sin\theta_1 =\sqrt{\frac{m_u}{nm_c}}\label{eq:thpred}
\end{eqnarray}
In the basis where the up-quark mass matrix is diagonal,
the renormalized down-quark and charged lepton mass matrices are given by
\begin{eqnarray}
m_D^{ren} &\approx & \gamma_D I_{\xi}{\it K^{\dagger}}
m_D{\it K}\label{eq:rdown}\\
m_E^{ren} &\approx & \gamma_E {\it K^{\dagger}}m_E{\it K} \label{eq:relec}
\end{eqnarray}
where $I_{\xi}=Diagonal(1,1,\xi)$,
and $m_{D,E}$, are given in Eqs(\ref{eq:downq}) and (\ref{eq:elec}).
In order to make definite predictions in the fermion mass sector, we choose
the lepton masses as inputs and express the arbitrary parameters
$a,b,\lambda_{\tau}(\equiv \lambda_b) $ in $m_E$ in terms of $m_e,m_{\mu}$,
and $m_{\tau}$. Substituting into $m_D$ and diagonalizing $m_D^{ren}$, we
obtain\cite{lv}
\begin{eqnarray}
m_d\approx  6.3\times (\frac{\eta_d}{2})MeV \\
m_s\approx 153 \times (\frac{\eta_s}{2})MeV \\
m_b \approx \eta_b \frac{\gamma_D}{\gamma_E}\xi m_{\tau}\label{eq:bottom}
\end{eqnarray}
Since $\eta_d\approx \eta_s\approx 2$, the predictions for the light quarks
$m_{d,s}$ are within the expected range\cite{qcd}.
Now, in order to make a prediction for the bottom quark, we need to know
the value of $\xi$, but the latter depends on the top quark
coupling at the GUT
scale  as well as on the top-mass, through Eq.(\ref{eq:ksi}).
 We can use, however,
Eq.(\ref{eq:bottom}), to predict the range of the top mass.
Thus, using the available limits
on the bottom mass, $4.15 \le m_b\le 4.35 GeV$, and $\eta_b\approx 1.4$,
 we can obtain the
following range for $m_{top}$
\begin{eqnarray}
125GeV\le m_{top}  \le 170 GeV\label{eq:toprange}
\end{eqnarray}

As we work in a diagonal basis for the up-quark mass matrix,
the CKM - matrix
can be found by diagonalizing $m_D^{ren}$ in Eq(\ref{eq:rdown}).
Let us denote with $U_{\theta_2}$ the matrix which diagonalizes the
$m_D(M_{G})$: \begin{eqnarray}
U_{\theta_2}&=&\left(\begin{array}{ccc}
cos\theta_2e^{-\imath\phi}&-sin\theta_2&0\\sin\theta_2&cos\theta_2e^
{-\imath\phi}&0 \\0&0&1\end{array} \right) \label{eq:ckm}
\end{eqnarray}
Then the angle $\theta_2$,
 can be determined in terms of the lepton masses making
use of the relations between $m_D$ and $m_E$ entries at the GUT scale. We find,
$sin\theta_2\approx \frac{m_e}{m_{\mu}} $. The CKM-matrix can now be expressed
to a good approximation by  $V_{CKM}\approx
U_{\theta_2}I_{\xi}K^{\dagger}I_{\xi}^{-1}U_{\theta_2}^{\dagger}$. In
particular,
 we get the following predictions for $V_{cd}$,$V_{ts}$ and $V_{td}$:
 \begin{eqnarray}
V_{cd}\approx  (-s_1c_2e^{\imath\phi}-s_2)/D_1\\
V_{ts}\approx  \xi s_1((n-1)s_1c_2+e^{\imath\phi}c_2)/D_3\\
V_{td}\approx  \xi s_1((n-1)s_1c_2-s_2)/D_3\label{eq:KMij}
\end{eqnarray}
where $c_i=cos\theta_i$ and $s_i=sin\theta_i$,$i=1,2$. As we have shown, both
angles are given in terms of well known low energy fermion masses, while from
the experimental range of $V_{ij}$ the parameter $n$ is constrained to be in
the range $3\le n \le 6$.
%%%%%%%
%%%%%%%%%%%%%%%%%%%%%%%%%%%%%%%%%%%%%%%%%
%%%%%%%%
\vglue 0.6cm
{\bf  3. Evolution of the scalar masses}

%%%%%%%%%%%%%%%%%%%%%%%%%%%%%%%%%%%%%%%%%%%%%%%%%

A particular role in flavour changing processes in supersymmetric
theories \cite{NLN}
 is played by the scalar masses. In this section, we are going
to present analytic formulae for the squark masses which are entering
the various processes to be examined in the next section.
After the renormalization group running,
the scalar masses, relevant to our discussion,
 are given in general by the following formula\cite{RAD,EKN,CK}
\begin{eqnarray}
{\tilde m}_i= \alpha_im^2_{3/2}+C_im^2_{1/2}\label{eq:mscal}
\end{eqnarray}
where $\alpha_i$ in general depend on
the K\"ahler function, but here we will assume
$\alpha_i\equiv \alpha_Q=\alpha_U=\alpha_E=1$
in the
minimal case where the scalars are in a flat manifold\cite{CFGP}.
$C_i(t)=C_Q,C_U,C_E$, are calculable coefficients
 which represent gauge corrections. Here, we are considering
cosmologically acceptable modes(CAM)\cite{EKN,CK}
with non-minimal number of
higgses.
In this case  their
numerical values at  $t\sim lnm_t$ can be found
from the approximatele formulae\footnote{In the
case of minimal models with three families and two higgses,
one finds\cite{CK}
$C_Q\approx 5.29,C_U\approx 4.84,C_E\approx .50.$}
\begin{eqnarray}
C_Q& \approx &ln\left(\prod_{j=1}^3 \left(\frac{
\alpha_{j,0}}{\alpha_{j}}\right)^{c_Q^j/b_j}\right)\nonumber \\
&\equiv & 2ln\gamma_Q \approx 2.32
\end{eqnarray}
and similarly for the other gauge coefficients
\begin{eqnarray}
C_U=2ln\gamma_U\approx 1.84,\,\, C_E=2ln\gamma_E\approx 0.62.
\end{eqnarray}
%%%%%%%%%%%%%%%%%%%%%%%%%%%%%%%%%%%%%%%%%%%%%%%%%
However, additional radiative corrections due to the heavy top quark,
 modify the above formulae for the top-squarks $m_{\tilde t_L},
m_{\tilde t_R}$, the  higgs mass $m_{\bar h}$, (where $\bar h$ is
 the higgs which gives masses to the up-quarks) and
the trilinear scalar coupling parameter $A$.
For later convenience, let us denote $m_{\tilde t_L}\equiv \tilde m_1$,
$m_{\tilde t_R}\equiv \tilde m_2$, and
$m_ {\bar h}\equiv \tilde m_3$.
Then, at any scale $t=ln{\mu}$, we can write\cite{CK}
 \begin{eqnarray}
\tilde m^2_n(t)=\alpha_nm_{3/2}^2+C_n(t)m_{1/2}^2
-n\Delta^2_A(t)-n\Delta^2_{\tilde m}(t)
\label{eq:YCn}
\end{eqnarray}
In the above,
\begin{eqnarray}
\Delta^2_A(t)+\Delta^2_{\tilde m}(t)=\int_{t}^{t_0}
\frac{\lambda_{top}^2(t^{\prime})}{8\pi^2}
\left(A^2(t^{\prime})+
\sum_{j=1}^3\tilde m^2_j(t^{\prime})\right)\,dt^{\prime}
\label{eq: Delta}
\end{eqnarray}
where with $\Delta_A(t)$ we have denoted the radiative corrections due to
the trilinear scalar coupling $A$ which in general depends on the scale
parameter $t=ln\mu$\cite{EKN}
\begin{eqnarray}
 A(t)=\frac{\alpha_0m_{3/2}+C_A(t)m_{1/2}}{1-12ln\xi}
\end{eqnarray}
 Thus, Eq(\ref{eq:YCn}), can be transformed to an integral equation
 which can be solved exactly\cite{GL}. Indeed, summing up all
the scalar masses containing the Yukawa corrections, we get
\begin{eqnarray}
\sum_{n=1}^3\tilde m^2_n(t)=
\sum_{n=1}^3\alpha_nm_{3/2}^2+
\sum_{n=1}^3C_n(t)m_{1/2}^2-\sum_{n=1}^3n\Delta_A^2(t)
-\sum_{n=1}^3n\Delta^2_{\tilde m}(t)
\label{eq: sumn}
\end{eqnarray}
If we further define,
\begin{eqnarray}
u(t)=\sum_{n=1}^3\tilde m^2_n(t)\nonumber\\
u_0(t)=\sum_{n=1}^3\alpha_nm_{3/2}^2+
\sum_{n=1}^3C_nm_{1/2}^2-\sum_{n=1}^3n\Delta_A^2(t)\label{eq: def}\\
C=-\frac{1}{8\pi^2}\sum_{n=1}^3n\nonumber
\end{eqnarray}
  Eq(\ref{eq: sumn}) can be cast in a standard integral form
\begin{eqnarray}
u(t)=
u_0(t)+C\int_{t}^{t_0}
{dt^{\prime}\lambda_{top}^2(t^{\prime})u(t^{\prime})}\label{eq: ut}
\end{eqnarray}
%%%%%%%%%%%%%%%%%%%%%%
Solving the above integral equation, we find that
 the scalar masses are given by\cite{GL}
\begin{eqnarray}
\tilde m ^2 _n(t)&=&\alpha_nm_{3/2}^2+C_n(t)m_{1/2}^2
{}-{}n\delta_m^2(t){}-{}n\delta_A^2(t)
\label{eq: smn}
\end{eqnarray}
where
\begin{eqnarray}
\delta_m^2(t)=\left(\frac{m_t}{2\pi
 \upsilon\gamma_Q sin\beta}\right)^2
\times (3m_{3/2}^2I+m_{1/2}^2J)\label{eq:dm1}
\end{eqnarray}
and,
%%%%%%%%%%%%%%%%%%%%%
\begin{eqnarray}
\delta_A^2(t)&=&\Delta_A^2(t)
-\frac{3}{2}\left(\frac{m_t}{2\pi
 \upsilon\gamma_Q sin\beta}\right)^2E_A^2(t)
\end{eqnarray}
where $I,J$ and $E_A^2$, are integrals containing functions of
gauge couplings, i.e.
\begin{eqnarray}
I&=&\int_{t}^{t_0}\,dt^{\prime}  \gamma_Q^2(t^{\prime})\\
J&=&\int_{t}^{t_0}\,dt^{\prime}  \gamma_Q^2(t
^{\prime})C(t^{\prime} )\\
E_A^2&=&\int_{t}^{t_0}\,dt^{\prime}  \gamma_Q^2(t
^{\prime})\Delta_A^2(t^{\prime} )
\label{eq: IJE}
\end{eqnarray}
with $C(t)\equiv \sum_{n=1}^3C_n(t)$.

In the above expressions use has been made of
$\lambda_{top}^2(t)$ from Eq(\ref{eq:ltop}),
and the top-mass  formula
$m_t=\lambda_{top}\frac{\upsilon}{\sqrt 2}sin\beta$
where $\upsilon =246GeV$

We note in passing, that contributions from Yukawa radiative corrections
to the scalar masses, are always negative. This fact is of crusial
imporance, since it allows the possibility of breaking the $SU(2)\times
U(1)$ symmetry radiatively.\cite{IR,RAD,EKN}
Indeed, substituting numerical values into Eq(
\ref{eq: smn})  gives a negative $(mass)^2$ for the mass
$\tilde m_3^2\equiv  m_{\bar h}^2$, as long as $m_t\ge 155sin\beta
GeV$\cite{GL}. This bound is consistent with the ansatz discussed in the
previous section, since the top mass is always given by the approximate
relation, $m_t\approx 180sin\beta GeV$.

%%%%%%%%%%%%%%%%%%%%%%%%%%%%%%%%%%%%%%%%%%%%%%%%%%%%%%%%%%%%%%%%%%%%%%

\vglue 0.6cm
{\bf  4. Flavour Changing Neutral Currents}
 \vglue 0.4cm

It has been pointed out by several authors\cite{Dug,Mas,GMas,IL,Hag} that
flavour changing  neutral currents
may be enhanced  in various extentions of the Minimal Supersymmetric
Standard Model (MSSM).
In particular, renormalization group effects are a significant source of
 flavour violation
in SUSY GUTs. Indeed, the running of the fermion masses from the GUT scale
 down to the low
energies, is different from that of their supersymmetric partners. Thus,
fermion
and s-fermion mass matrices of the same type, cannot be simultaneously
diagonalized. This
fact leads to off-diagonal couplings with neutral gauginos and therefore
 to flavour changing neutral currents.
In particular, if we assume that we are in a basis where $m_Dm_D^{\dagger}$
is diagonal,
then the supersymmetric matrix $m_{\tilde D}m_{\tilde D^*}$, receives
 contributions
from the  $\lambda_UQHu^c$ coupling of the superpotential, through the
renormalization group running, leading to a non-diagonal correction of
 the form
\begin{eqnarray}
\delta (m_{\tilde D}m_{\tilde D^*})\approx cm_Qm_Q^{\dagger}
\label{eq:deltam}
\end{eqnarray}
The parameter $c$ is determined by solving the
 renormalization group equations in the spirit of the previous section.
In the limit where $\lambda_{top}\gg \lambda_{u,c}$, we may use the results
of  the previous section, and write the above contribution as follows
\begin{eqnarray}
\delta (m_{\tilde D}m_{\tilde D^*})_{ij}\approx
 -\left(\frac{g_2}{4\pi m_W\gamma_Q
sin\beta}\right)^2V_{ti}^{*}m_t^2V_{tj} \times
(3m_{3/2}^2I+m_{1/2}^2J)\label{eq:deltam1} \end{eqnarray}
The parameter $c$ can now be extracted by direct comparison of the
 Eqs(\ref{eq:deltam},\ref{eq:deltam1}).

It is convenient in the following, to define the flavour violating
quantities
\begin{eqnarray}
\frac {\delta \tilde m_{ij}^2}{\tilde m^2}\equiv
\frac {\delta (m_{\tilde D}m_{\tilde D^*})_{ij}}{m_{\tilde t_L}^2}
\label{eq:fcq}
\end{eqnarray}
and parametrize the various flavour changing processes in terms of them.
 Thus all the dependence on the quark masses and the CKM- mixing angles
is contained in Eq(\ref{eq:fcq}). Furthermore,
it is possible to eliminate the $m_t$- dependence  by observing that in the
particular ansatz,
$m_t$ is given by $m_t\approx 180sin\beta GeV $, to a
pretty good approximation.
Now, the quantities entering  the various flavour changing processes,
 can be writen in the following
way
\begin{eqnarray}
\frac {\delta \tilde m_{ij}^2}{\tilde m^2}=-\frac{\alpha_2}{4\pi}
\left(\frac{180GeV}{m_W\gamma_Q}\right)^2
(V^{*}_{ti}V_{tj})\frac{3m^2_{3/2}I+m^2_{1/2}J}
{\alpha_Qm^2_{3/2}+C_Qm^2_{1/2}-\delta_m^2}\label{eq:del180}
\end{eqnarray}
where  the squark mass $\tilde m$, has been substituted from the Eq.(
\ref{eq: smn}). In contrast to naive estimations\cite{GMas,Hag}, we should
 notice in the above formula the appearance of the additive term
proportional to $m_{1/2}$, which is comparable to the term proportional
to $m_{3/2}$ ($J\sim 5I)$ and arises from the integration of the gauge
part.
The values of $I,J$ integrals of the previous section,
are found to be $I\approx 113$ and  $ J\approx 580$. (
note also that $\delta_A^2 $
contributions have been ignored, since it is found that
 $\delta_A^2\ll\delta_m^2$). Thus, the above exact calculation of the
Yukawa type corrections to the scalars, results
 at least to an enhancement by
a factor of $2$ of the flavour violating quantities, compared to
previous estimations. Finally, we should remark that
it is adequate for our purposes in Eq(\ref{eq:del180})
to include only  corrections proportional to $(V^{*}_{ti}m_t^2V_{tj})$
terms since $m_t\gg m_{u,c}. $(However, note that
inclusion of the first two generations should involve different integral
factors $I,J\rightarrow I^{\prime} ,J^{\prime} $.)
If we define, $\tilde x=\frac{m_{3/2}}{m_{1/2}}$
and substitute all the known values of the various quantities in this last
expression, we finally get
\begin{eqnarray}
\frac {\delta \tilde m_{ij}^2}{\tilde m^2}\approx 0.45\times
V^{*}_{ti}V_{tj}\frac{\tilde x^2+1.42}
{0.55\tilde x^2+1.42}\label{eq: fcx}
\end{eqnarray}
Since the CKM angles
in this last expression are predicted by the fermion mass matrix ansatz,
the flavour violating quantities
 depend basically only on the ratio $\tilde x=\frac{m_{3/2}}{m_{1/2}}$,
 thus all the uncertainties arising from the top range the higgs vev ratio
etc, have been minimized.
In what follows, we are going to discuss  the implications of the above
renormalization group effects on certain flavour changing processes.

{\it {\bf i) $K^0-\bar K^0$ system.}}
If $L_{\Delta S}^{eff}$ represents the $\Delta  S=2$ effective lagrangian,
then the $K_L - K_S$  mass difference can be expressed as follows
\begin{eqnarray}
\Delta K = \frac{1}{m_K}<K^0|-L_{\Delta S}^{eff}|\bar K^0>
\end{eqnarray}
It can be shown, that the dominant supersymmetric contribution to
$K^0-\bar K^0$ mixing arises from the  $\delta \tilde m_{LL}^2$ flavour
changing quantities, which involve only $(V-A)^2$-type operators.
Therefore, let us assume only $LL$ operators for simplicity;
 The relevant matrix element in this case is
\begin{eqnarray}
<K^0|[\bar d_{\alpha}\gamma_{\mu}\frac{1-\gamma_5}{2}s_{\alpha}]
[\bar d_{\beta}\gamma^{\mu}\frac{1-\gamma_5}{2}s_{\beta}]|\bar K^0>=
\frac{2}{3}f_K^2m_KB_K^{\alpha}
\end{eqnarray}
in the case of $K^0-\bar K^0$ mass difference the relevant mass
parameter is $\delta \tilde m_{\bar d s}^2\equiv \delta \tilde m_{12}^2$,
thus we have
%%%%%%%%%%%%%%   %%%%%%%%%%%%  %%%%%%%%
\begin{eqnarray}
\frac {\delta \tilde m_{12}^2}{\tilde m^2}\approx 0.45\times
V^{*}_{t1}V_{t2}\frac{\tilde x^2+1.42}
{0.55\tilde x^2+1.42}
\label{eq:delm}
\end{eqnarray}
In this case we obtain a mass difference\cite{GMas,Hag}
\begin{eqnarray}
\Delta M_K \approx  \frac {\alpha_s^2}{216\tilde m^2}
(\frac{2}{3}f_K^2m_KB_K^{\alpha})
(\frac {\delta \tilde m_{12}^2}{\tilde
m^2})^2\left[-66G(r)-24r^2F(r)\right] \label{eq:delm12}
\end{eqnarray}
In this formula $a_s$ is the strong gauge coupling, $f_K\approx 165MeV$ is
the K-meson decay constant, $B_K^{\alpha}\le 1$ is the
``fudge factor''and $m_K
\approx 497.671MeV$
The functions $G(r)$ and
$F(r)$ are given by
 \begin{eqnarray}
G(r)=\frac{1}{3(1-r^2)^5}(r^6+9r^4-9r^2-1-12r^2(1+r^2)lnr)\label{eq:Gr}\\
F(r)=\frac{1}{6(1-r^2)^5}(-r^6+9r^4+9r^2-17-12(1+3r^2)lnr)\label{eq:Fr}
\end{eqnarray}
where
\begin{eqnarray}
r&\equiv &\frac {m_{\tilde g}}{\tilde m}\nonumber\\
{}&\approx &\frac{\alpha_s}{\alpha_G}\frac{1}{0.55\tilde x^2
+1.42}\label{eq:ro} \end{eqnarray}
This contribution to $\Delta m_K$ of course should be smaller than the
experimental value
\begin{eqnarray}
(\Delta M_K)^{exp} \approx  3.522\times 10^{-15}GeV\label{eq:m12exp}
\end{eqnarray}
Taking the maximum value for $B_K^{\dagger}=1$, $a_s=.11$
in Eq(\ref{eq:delm12})
and making use of  (\ref{eq:m12exp}), we can convert the above constraint
to a limit on the gravitino mass $m_{1/2}$ as a function of the ratio
 $\tilde x=\frac{m_{3/2}}{m_{1/2}}$  i.e.
\begin{eqnarray}
m_{1/2}>7.63\times (10^3V_{t1}^{*}V_{t2}) \frac {\tilde x^2+1.42}
{(0.55\tilde x^2+1.42)^{3/2}}f(r)GeV
\label{eq:m1/2k}
\end{eqnarray}
or, using $m_{1/2}=\frac{m_{3/2}}{\tilde x}$, we can interpret it
 as a bound on $m_{3/2}$, i.e.
\begin{eqnarray}
m_{3/2}>7.63\times (10^3V_{t1}^{*}V_{t2}) \frac {\tilde x(\tilde x^2+1.42)}
{(0.55\tilde x^2+1.42)^{3/2}}f(r)GeV
\label{eq:m0k}
\end{eqnarray}
These bounds depend crucially on the function $f(r)$.
This function equals unity for $r=1$, increases rapidly as $r$ decreases,
while decreases for $r\ge 1$. Thus, from Eq(\ref{eq:ro}) it is clear that
there is an enhancement of the $m_{3/2}$ bound for $\tilde x\gg 1$.
There is a significant suppression however
in this case due to the CKM elements. In the fermion mass texture discussed
in section 2, for the allowed range of the top-quark mass, the
predicted range is $(10^3\times V_{t1}^{*}V_{t2}) \approx 0.368-0.516$.
Thus, the  $m_{3/2}$, $m_{1/2}$ lower
bounds from $K^0-\bar K^0$ are relatively small. Constraints from
 $B_d-\bar B_d$ mixing can also be treated similarly .

More stringent lower bounds on the above parameters can be obtained from
the CP-impurity parameter $\epsilon _K$. Experimentally we know that
\begin{eqnarray}
\frac{Im M_{12}}{Re M_{12}}\approx 2\sqrt{2}\epsilon _K \approx
6.5\times10^{-3}\label{eq:eps}
 \end{eqnarray}
Supersymmetric contributions should be smaller than the above value.
A straightforward calculation shows that\cite{Hag}
\begin{eqnarray}
\mid\frac {\delta \tilde m_{12}^2}{\tilde m^2}\mid\le
\frac{18\tilde m}{\alpha_sf_K}\sqrt{\frac{
2\sqrt{2}\epsilon_K(\Delta M_K)}{sin(2\phi)f(r)m_K}}
 \end{eqnarray}
In the present case one obtains one order of magnitude larger
bound than that from the $K^0-\bar K^0$ system, i.e.

\begin{eqnarray}
m_{1/2}\geq 67\times (10^3V_{t1}^{*}V_{t2})
 \frac {\tilde x^2+1.42}{(\tilde 0.55x^2+1.42)^{3/2}}f(r)GeV
\label{eq:m0keps}
\end{eqnarray}
which can also be converted to an $m_{3/2}$ bound as previously.

The obtained bounds are presented in figures (1-4).
In figure 1, we plot the resulting bounds on the $m\equiv m_{1/2}$ for the
higher as well as the lower predicted value of the product
$V_{t1}^{*}V_{t2}$, while in figure 2 the corresponding bounds are
for the $m_0\equiv m_{3/2}$ scalar mass parameter, both,
as a function of their ratio  $\tilde x=\frac{m_{3/2}}{m_{1/2}}$.
 For small $\tilde x$ ratios the bounds
are also small. However, as their ratio increases, $m_{3/2}$ approaches
asymptotically its maximum value which lies between $m_{3/2}\approx
(300-400)GeV$, depending on the precise value of the product
$V_{t1}^{*}V_{t2}$. $m_{1/2}$ bounds are significantly smaller, with a
maximum
 $\approx (40-60)GeV$, when $\tilde x \approx 3$.
Figures 1 and 2, should be considered in conjunction with
 figure 3, which shows that the allowed region in the parameter space
$(m_{3/2},m_{1/2})$. The exterior space of the two curves represents the
allowed $(m_{3/2},m_{1/2})$ pairs for the two extreme $V_{t1}^{*}V_{t2}$
values discussed previously.

Interesting bounds may also be obtained for the low
energy scalar masses. Thus for example, the $m_{\tilde t_L}$ squark mass
bound is presented in figure 4, in the parameter space $m_{3/2},m_{1/2}$.
 Of course, only the part of the region
which satisfies the constraints from figure 3 is accepted.
Obviously, the intersection of the two figures allows only
maximum values  $m_{\tilde t_L}\ge (200-300)GeV$.
Similarly, bounds on the other scalar masses are obtained from the
formula (\ref{eq:mscal})

{\it ii) $b\rightarrow s+\gamma $}. Recently there has been a revived
interest{\cite{BG,LNPZ}} on this reaction, prompted by the CLEO
bound\cite{cleo}  ($BR(b\rightarrow s+\gamma) < 8.4\times 10^{-4}$).

The general expression for $b\rightarrow s+\gamma$ can be given by
 \cite{BG}
\begin{equation}
\frac{B(b\rightarrow s\gamma)}{ B(b\rightarrow
 ce\bar\nu)}=\frac{6\alpha}{\pi}
\frac{\left[\eta^{16/23}A_\gamma
+\frac{8}{3}(\eta^{14/23}-\eta^{16/23})A_g+C\right]^2}{
I(m_c/m_b)\left[1-\frac{2}{3\pi}\alpha_s(m_b)f(m_c/m_b)\right]},
\end{equation}
where  $I$ is the phase-space factor
$I(x)=1-8x^2+8x^6-x^8-24x^4\ln x$,
$\eta=\alpha_s(M_Z)/\alpha_s(m_b)$, and $f(m_c/m_b)=2.41$ the QCD
correction factor for the semileptonic decay. $A_\gamma,A_g$ are
coefficients\cite{BG} of the effective $bs\gamma$ and $bsg$ penguin
operators. It can be shown\cite{LNPZ}, that these contributions are
 not far away from the reported CLEO bound and
 in some cases they exclude a significant fraction of
the region in the $(m_t,\tan\beta)$ parameter space.

It is interesting therefore to discuss also
bounds arising from the $b\rightarrow s+\gamma $ decay due to
renormalization group effects.
Since $LL$-insertions are found most important, we
consider again diagrams which
contribute only to the $LL$-case\cite{GMas,Hag}. We may demand that this
contribution is smaller than the experimentally determined range
($(0.6-5.5)\times 10^{-4}$\cite{bsg}), to obtain
\begin{eqnarray}
m_{1/2}>(13-20)\times 10\sqrt{V_{t2}^{*}V_{t3}}
\frac{\sqrt{|(\tilde x^2+1.42)R(r)|}}{0.55\tilde x^2+1.42}
\label{eq:mbsg}
\end{eqnarray}
with
\begin{eqnarray}
R(r)=\frac{10}{3}
\frac{1}{(1-r^2)^5}(17r^6-9r^4-9r^2+1-12r^4(3+r^2)lnr)\label{eq:Rr}
\end{eqnarray}
We note again here that the  dependence on $m_t$ and $sin\beta$,
has been rotated away, thus there remains only the dependence
 on the CKM mixing angles and the  ratio $\tilde x$. Although
however, the CKM matrix elements are bigger ($V_{t2}^{*}V_{t3}
\approx (3.4-4.3)\times 10^{-2}$) than those involved
in the CP-impurity parameter calculations, the bounds
put by this reaction on the supersymmetric sparticle spectrum are
significantly lower.
\\

%%%%%%%%%%%%%%%%%%%%%%%%% CONCLUSIONS %%%%%%%%%%%%%%%%%%%%%%%%%%%%%%%%%%%
{\bf 5. Conclusions}

In this work, we have discussed  flavour changing neutral current
 reactions in the
context of predictive supersymmetric grand unified theories, taking into
account renormalization group effects. We have used a  particular fermion
mass matrix ansatz to predict the flavour  changing quantities
which depend on fermion masses and mixing angles, in order to minimize the
uncertainties in the ampitudes of the various flavour changing processes.
We have found that all the uncertainties arising from the Yukawa sector
(masses and mixing angles and the higgs ratio), for any flavour changing
reaction  $i\rightarrow j$ can be reduced down to a simple product only
of the CKM mixing angles, $V_{ti}^{*}V_{tj}$, which is also constrained
from the particular mass matrix ansatz.  Furthermore,
an exact calculation of the Yukawa corrections
to the scalar masses in the limit where
$\lambda_{top}\gg \lambda_{u,c,b}$, has resulted to an enhancement by
a factor of 2 with respect to previous estimations, of the flavour changing
quantities involved in the  various reactions.
The above results reconciled with experimental bounds on flavour changing
reactions have been used then to  put limits on the supersymmetric GUT
parameters $m_{3/2}$, $m_{1/2}$ as well as the low energy scalar masses.
It is found that the most stringent bounds are put by the CP-impurity
parameter $\epsilon_K$. In particular, there is a maximum lower bound
for $m_{3/2}\ge 300GeV$, as $m_{1/2}\rightarrow 0$.
 These bounds are also converted to
bounds for the various scalar masses of the low energy theory, since the
latter  can be expressed in terms of the former, as they are related by the
renormalization group equations.

%%%%%%%%%%%%%%%%%%%%%%%%%%%%%%%%%%%%%%%%%%%%%%%%%%%%%%%%%%%%%%%%%%%%%

%%%%%%%%%%%%%%%%%%%%%%%%%%%%%%%%%%%%%%%%%%%%%%%%%%%%%%%%%%%%%%%%%%%%%
\newpage

{\bf Figure Captions}
\vglue 0.4cm
{\bf Figure 1}. Lower bounds on $m\equiv m_{1/2}$ scalar mass parameter
as a function of the ratio $\tilde x =m_{3/2}/m_{1/2}$. The lower curve
corresponds to $V_{td}^{*}V_{ts}=3.67\times 10^{-4}$, while the upper
one to $V_{td}^{*}V_{ts}=5.16\times 10^{-4}$.
\vglue 0.2cm
{\bf Figure 2}. Lower limits for $m_0\equiv m_{3/2}$ SUSY scalar mass
parameter, for the two cases of figure 1.
\vglue 0.2cm
{\bf Figure 3}. Exluded (interior), and allowed (exterior) $m_{3/2},
 m_{1/2}$ regions from $\epsilon_K$ experimental bounds, for the
two values of $V_{td}^{*}V_{ts}$.
\vglue 0.2cm
{\bf Figure 4}. Lower bounds of $m_{\tilde t_L}$ squark mass in the
parameter space $m_{3/2}, m_{1/2}$. The part of the ($m_{3/2},
m_{1/2}$)$-$plane allowed by the two curves in figure 3, leads to
$m_{\tilde t_L}\ge (200, 300) GeV$ respectively.
\end{document}